\documentclass[12pt]{article}

\usepackage[doublespacing]{setspace}

\usepackage{amsmath}
\usepackage{graphicx}
\usepackage{geometry}
\usepackage[applemac]{inputenc}

\usepackage{natbib}

\usepackage[tight]{subfigure}
\ifx\pdfoutput\@undefined\usepackage[usenames,dvips]{color}
\else\usepackage[usenames,dvipsnames]{color}
\IfFileExists{pdfcolmk.sty}{\usepackage{pdfcolmk}}{} 
\fi
\usepackage[plainpages=false,pdfpagelabels,pagebackref=false,naturalnames=true,hyperindex=true,pdftitle={},pdfauthor={Carlos Gershenson}]{hyperref}
\hypersetup{colorlinks=true,
urlcolor=Cerulean,linkcolor=BrickRed,citecolor=RoyalBlue,
  pdfpagemode=None,
  pdfstartview=FitH}
\usepackage[all]{hypcap}

\begin{document}

\title{Complexity and Information: \\Measuring Emergence, Self-organization,\\ and Homeostasis at Multiple Scales}
\author{Carlos Gershenson$^{1,2}$ \& Nelson Fern\'{a}ndez$^{3,4}$ \\
$^{1}$ Departamento de Ciencias de la Computaci\'on\\
Instituto de Investigaciones en Matem\'aticas Aplicadas y en Sistemas \\
Universidad Nacional Aut\'onoma de M\'exico\\
A.P. 20-726, 01000 M\'exico D.F. M\'exico\\
\href{mailto:cgg@unam.mx}{cgg@unam.mx} \
\url{http://turing.iimas.unam.mx/~cgg} \\
$^{2}$ Centro de Ciencias de la Complejidad \\
Universidad Nacional Aut\'onoma de M\'exico\\
$^{3}$ Laboratorio de Hidroinform\'{a}tica, Facultad de Ciencias B\'{a}sicas\\
Univesidad de Pamplona, Colombia\\
\url{http://unipamplona.academia.edu/NelsonFernandez}\\
$^{4}$ Centro de Micro-electr\'onica y Sistemas Distribuidos,\\
Universidad de los Andes, M\'erida, Venezuela
}
\maketitle

\clearpage

\begin{abstract}
Concepts used in the scientific study of complex systems have become so widespread that their use and abuse has led to ambiguity and confusion in their meaning. In this paper we use information theory to provide abstract and concise measures of complexity, emergence, self-organization, and homeostasis. The purpose is to clarify the meaning of these concepts with the aid of the proposed formal measures. In a simplified version of the measures (focusing on the information produced by a system), emergence becomes the opposite of self-organization, while complexity represents their balance. \textcolor{black}{Homeostasis can be seen as a measure of the stability of the system.} We use computational experiments on random Boolean networks and elementary cellular automata to illustrate our measures at multiple scales. 

\begin{center}
{\bf Nontechnical Abstract}
\end{center}

There several measures and definitions of complexity, emergence, self-organization, and homeostasis. This has led to confusion reflected on the carefree use of these concepts. \textcolor{black}{We provide precise definitions based on information theory, attempting to clarify the meaning of these concepts} with simple but formal measures. We illustrate the measures with experiments on discrete dynamical systems.

\end{abstract}


\section{Introduction}

In recent decades, the scientific study of complex systems has increased our understanding of a broad range of phenomena and produced several useful tools~\citep{Bar-Yam1997,Mitchell:2009}. Concepts used to characterize complex systems---such as emergence, adaptivity, self-organization, and complexity itself---have been used in different contexts with different meanings~\citep{NicolisPrigogine1977,Haken1988,Holland1995,Schweitzer1997,Wolfram:2002,Newman:2003,Schweitzer2003,Chaitin:2004,batty2005cities,Morin2006,ProkopenkoEtAl2007}. 

The diversity---and sometimes ambiguity---of notions, definitions and measures of complexity~\citep{Edmonds:1999} and related concepts has not only hampered the path to an elusive unified framework, but also induced the abuse of the terminology in non-scientific discourses, leading in many cases to the confusion of the general public and the contempt of established disciplines. 

Joining the effort of the community to bring clarity and agreement to these essential questions, we propose general measures of emergence, self-organization, homeostasis, and complexity based on information theory. The measures are general enough so that several previously proposed definitions can be seen as particular cases of them. While being formal and precise, the proposed measures are simple enough so as to clarify the studied concepts to people without an extensive mathematical background.

In the next section, we present a brief review of information theory and how it has been applied to complexity, including the concepts of emergence, self-organization, and homeostasis. In Section \ref{sec:proposal}, we present our proposal of abstract measures and extend them to multiple scales. Section \ref{sec:experiments} describes our experiments and results with random Boolean networks and elementary cellular automata, which illustrate \textcolor{black}{the} proposed measures. This is followed by a discussion in Section \ref{sec:discussion}. The paper closes with proposals for future work and conclusions.

\section{Information Theory and Complexity}

\begin{figure}[htbp]
\begin{center}
  \includegraphics[width=0.5\textwidth]{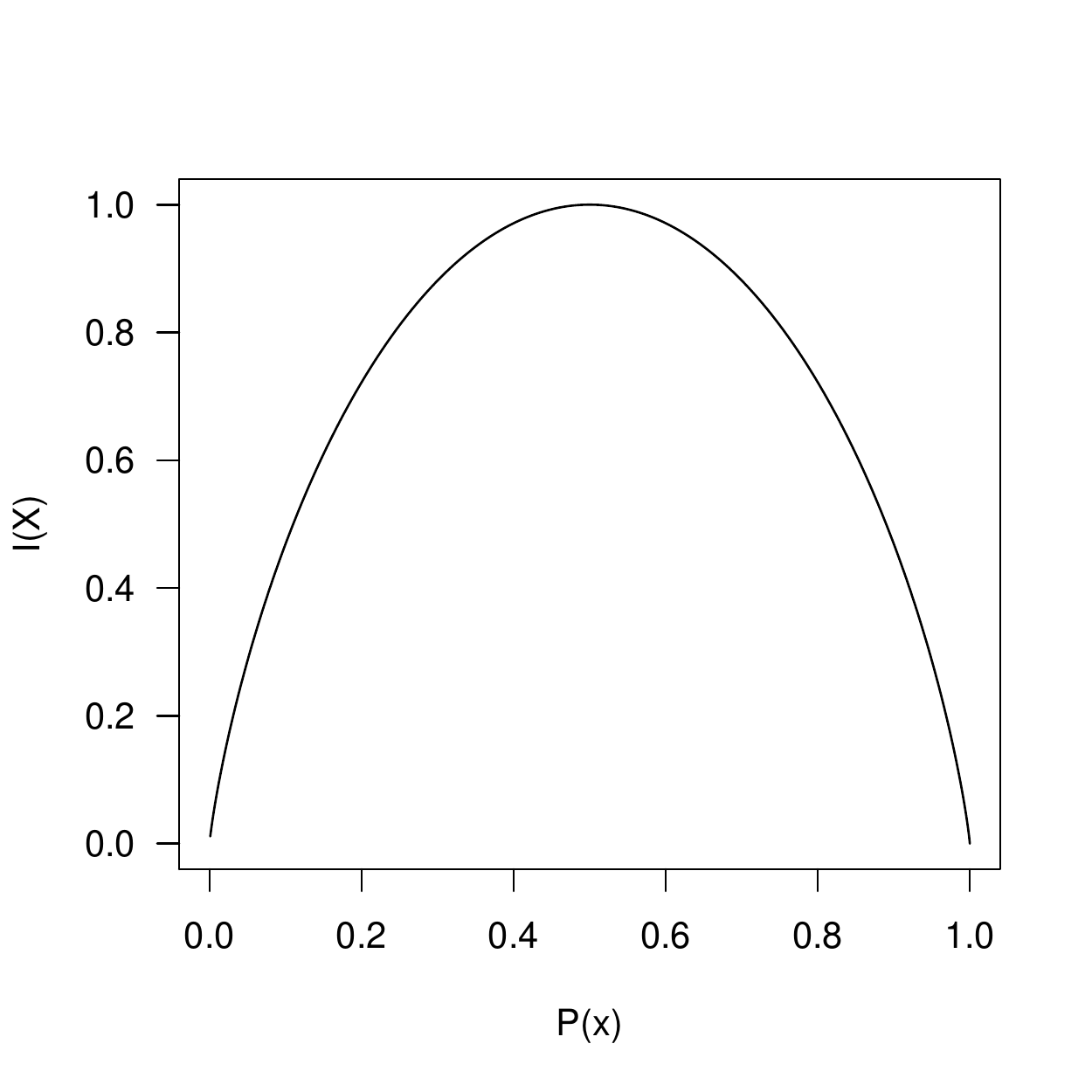}\\
\caption{Shannon's Information $I(X)$ of a binary string $X$ for different probabilities $P(x)$. Note that $P(0)=1-P(1)$.}
\label{fig:entropy}
\end{center}
\end{figure}

Information can be seen as a quantifiable pattern. Shannon~\citeyearpar{Shannon1948} studied it formally in the context of telecommunication. He was interested on how reliable communication could take place with unreliable components~\citep{gleick2011information}. \textcolor{black}{Given a string $X$, composed by  a sequence of values $x$ which follow a probability distribution $P(x)$, information (according to Shannon) is defined as: }

\begin{equation}
I=-\sum{P(x) \log P(x)}.
\label{eq:I}
\end{equation}
For binary strings, the logarithm is usually taken with base two. For example, if the probability of receiving ones is maximal ($P(1)=1$) and the probability of receiving zeros is minimal ($P(0)=0$), the information is minimal, i.e. $I=0$, since we know beforehand that the future value of $x$ will be $1$. There is no information because future values of $x$ do not add novel information, i.e. the values are known beforehand. If we have no knowledge about the future value of $x$, as with a fair coin toss, then $P(0)=P(1)=0.5$. In this case, information will be maximal, i.e. $I=1$, because a future observation will give us all the relevant information, which is also independent of previous values. Equation \ref{eq:I} is plotted in Figure \ref{fig:entropy}. Shannon information can be seen also as a measure of uncertainty. If there is absolute certainty about the future of $x$, be it zero ($P(0)=1$) or one ($P(1)=1$), then the information received will be zero. If there is no certainty due to the probability distribution ($P(0)=P(1)=0.5$), then the information received will be maximal.
\textcolor{black}{The unit of information is the bit. One bit represents the information gained when a binary random variable becomes known. If no information is gained, then $I=0$ bits. }

Prokopenko et al.~\citeyearpar{ProkopenkoEtAl2007} have produced a primer where proposals of measures of complexity, emergence, and self-organization based on information theory are reviewed. Their primer is much more extensive and detailed than this section.

\subsection{Complexity}

There are dozens of notions, definitions, and measures of complexity~\citep{Edmonds:1999,lloyd2001measures}. Etymologically, the term complexity comes from the Latin \emph{plexus}, which means interwoven. In other words, something complex is difficult to separate. This can be said to be because of relevant interactions between components~\citep{Gershenson:2011e}. However, when an observation is made in terms of information, a relevant question is ``how complex is this information?". In this case, interactions may be hidden from the observation, in many cases simply a string of bits or characters, without direct reference to the process that produced it. For this purpose, complexity can be seen as the amount of information required to describe a phenomenon at a particular scale~\citep{BarYam2004}. If more information is required, then the complexity can be said to be higher. However, the scale of observation is very important. For example, the information of the string  '1010101010' is maximal at a 1-bit scale (base 2), while it is minimal at a 2-bit scale (base 4) '22222'. 

Several measures of complexity assume that highly ordered strings ($I\approx0$) are not complex, while strings with a high information value ($I\approx1$) have a high complexity. In some cases, this makes sense, but in others, it implies that random or chaotic systems have the highest complexity. Several other measures---such as effective complexity~\citep{GellMann:1996}, Langton's $\lambda$~\citep{Langton1990} and Wuensche's $Z$~\citep{Wuensche1999}---see complexity as a balance between ordered and chaotic dynamics. In this direction, the measure proposed by L\'opez-Ruiz et al.~\citeyearpar{LopezRuiz:1995} is interesting, since they measure complexity as the multiplication of Shannon's information and disequilibrium, in the sense of ``far from thermodynamical equilibrium", where entropy is maximal~\citep{NicolisPrigogine1977}. In this way, highly ordered systems, such as crystals, will have and information close to zero, a high disequilibrium, and thus a low complexity. On the other hand, highly chaotic systems, such as gases, will have a high information, a disequilibrium close to zero (since they are close to thermodynamical equilibrium), and also a low complexity. High complexities are achieved for balanced values of information and disequilibrium.

\subsection{Emergence}

The concept of emergence has been studied for centuries. Intuitively, emergence refers to properties of a higher scale that are not present at the lower scale. For example, the color, conductivity, malleability of gold is not present in the properties of gold atoms~\citep{Anderson1972}. Emergence can be also seen as a change of description or as a metamodel~\citep{Heylighen1991b}. There have been several types of emergence proposed~\citep{Bedau:1997,BarYam:2004}, and there are problems that depend more on a philosophical perspective than on mathematics. 

Shalizi proposed a measure of efficiency of prediction based on information theory~\citep{Shalizi2001,ProkopenkoEtAl2007} which can be used to study which scale is more efficient for predicting the future. For example, it is useful to speak about emergent thermodynamics, since prediction is more efficient at a macro scale than at a micro scale, where information of individual atoms is taken into account.

In another measure of emergence, Holzer and De Meer~\citeyearpar{Holzer:2011}  compare the information at the system level (sum of all edges in a network) with the information of a lower level (a single edge). This measure gives a high value of emergence for systems with many dependencies (interdependent components) and a low value of emergence for systems with few dependencies (independent components).

\subsection{Self-organization}

Intuitively, self-organizing systems are those that produce a global pattern from the interactions of their components~\citep{GershensonDCSOS}. Classic examples include insect swarms, flocks of birds, and schools of fish~\citep{CamazineEtAl2003}. Independently of their mechanism, it can be said that self-organizing systems are those that increase their organization in time from their own internal dynamics. As Ashby~\citeyearpar{Ashby1947sos} noted, any dynamical system can be described as self-organizing, since they tend to attractors, and we only have to call those attractors ``organized" to call the system self-organizing. Thus, self-organization depends  partially on the purpose of the observer and the meaning she gives to the most probable states of a system. This does not imply that self-organization is purely subjective. Shannon information can be used to measure organization: ordered, organized strings have less information than chaotic, disorganized strings.

Gershenson \& Heylighen~\citeyearpar{GershensonHeylighen2003a} proposed to measure self-organization as the negative of the change of information $\Delta I$: if information is reduced, then self-organization occurs, while an increase of information implies self-disorganization. Nevertheless, it was shown that the same system can be considered as self-organizing or self-disorganizing, depending on the scale and on the partition of the state space~\citep{GershensonHeylighen2003a}.

\subsection{Homeostasis}

Cannon~\citeyearpar{Cannon:1932} defined homeostasis as the ability of an organism to maintain steady states of operation, in view of the internal and external changes. However, homeostasis does not imply an immobile or a stagnant state. Although some conditions may vary, the main properties remain relatively constant. Ashby~\citeyearpar{Ashby1947,Ashby:1960} later recognized that homeostasis corresponds to an adaptive reaction to maintain ``essential variables" within a range. Through adaptation, an organism develops the proper organization to function within a ``viability zone".  This viability zone is defined by the lower and upper bounds of the essential variables. In this sense, homeostasis is related with the capacity to remain within the viability zone of a system.

A dynamical system has a high homeostatic capacity if it is able to maintain its dynamics close to a certain state or states (attractors). When perturbations or environmental changes occur, the system adapts to face the changes within the viability zone, i.e. without the system ``breaking"~\citep{Ashby1947}. Homeostasis is also strongly related to robustness~\citep{Wagner2005,Jen2005}.

\section{An abstract proposal}
\label{sec:proposal}

Our proposal builds on previous work~\citep{Fernandez:2010,Fernandez:2012}. The main purpose of the measures presented below is to clarify the meaning of the concepts they represent, while being formal and precise. 

Information can be seen as a defined structure~\citep{Cohen2000,Cohen2006}, in other words, a \emph{pattern}. Any information has a pattern, let it be ordered, recursive, complex, or chaotic. 

One way of understanding emergence is as information at a higher scale that is not present at a lower scale. This can be generalized in terms of computation (information transformation~\citep{Gershenson:2007}). If we describe the dynamics of a system as a process, then we can define emergence as the novel information produced by that process which was not present beforehand. Emergence as a product of a change of scale is a particular case, since the change of scale can be seen as a process that transforms information.  If we are interested on how much information was produced by a process, we have to consider how much information was introduced. Thus, emergence $E$ can be formalized as:

\begin{equation}
E=\frac{I_{out}}{I_{in}},
\label{eq:E}
\end{equation}
where $I_{in}$ is the ``input information" (which can also be seen as an initial state or conditions) and $I_{out}$ is the ``output information", which can be seen as $I_{in}$ transformed by a computational process or function $f$, i.e. $I_{out}=f(I_{in})$. Emergence \textcolor{black}{is proportional to} the information \textcolor{black}{ratio} \emph{produced} by a process. \textcolor{black}{Since emergence is considered as a ratio, it is a unitless measure.}
If there is a random input, then $I_{in}=1$ can be assumed, so emergence is simplified to:

\begin{equation}
E=I_{out}.
\label{eq:Es}
\end{equation}

Concerning self-organization, we have already seen that this can be measured as a change of organization in time~\citep{GershensonHeylighen2003a}. Organization can be seen as the opposite of information, since high organization (order) is characterized by a low information and low organization (chaos) is characterized by a high information. In the context of information transformation, self-organization $S$ can be defined as:

\begin{equation}
S=I_{in}-I_{out}.
\label{eq:S}
\end{equation}
This implies that self-organization occurs ($S>0$) if the process reduces information, i.e. $I_{in} > I_{out}$. If the process generates more information, i.e. emergence occurs, then $S$ will be negative. \textcolor{black}{S measures information increase or reduction, so its units are also bits.} As with emergence, if a random input $I_{in}=1$ is assumed, then self-organization can be simplified to:

\begin{equation}
S=1-I_{out}.
\label{eq:Ss}
\end{equation}

It should be noted that $S$ represents how much order there is in a system, while $E$ represents how much variety there is. Following L\'opez-Ruiz et al.~\citeyearpar{LopezRuiz:1995}, we can define complexity $C$ as their multiplication:

\begin{equation}
C=E*S.
\label{eq:C}
\end{equation}

\begin{figure}[htbp]
\begin{center}
  \includegraphics[width=0.5\textwidth]{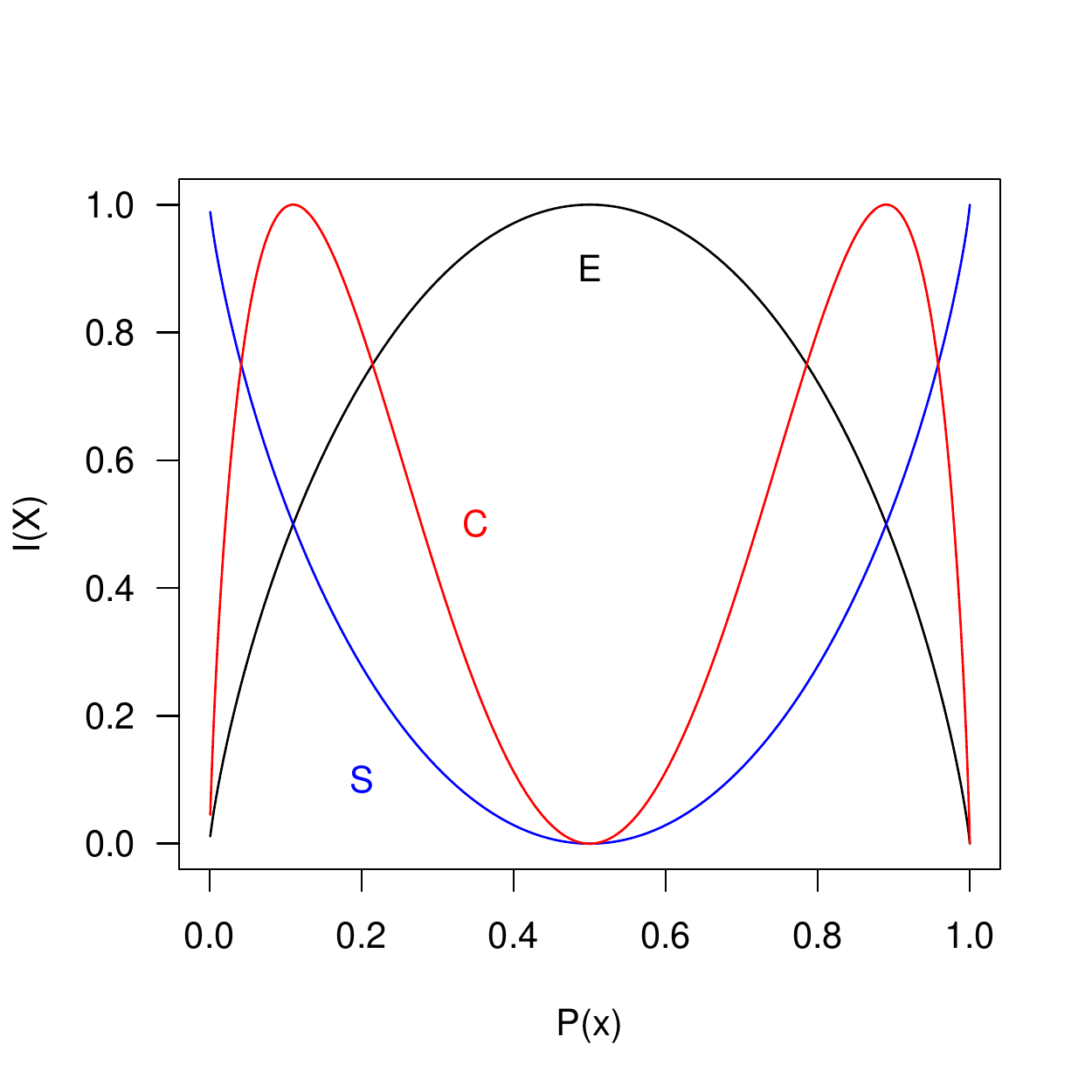}\\
\caption{Emergence $E$, self-organization $S$, and complexity $C$ of a binary string $X$ for different probabilities $P(x)$, considering random inputs ($I_{in}=1$).}
\label{fig:entropy2}
\end{center}
\end{figure}

Since a high $E$ implies a low $S$ and vice versa, a high $C$ can occur only when $E$ and $S$ are balanced. This is consistent with several notions of complexity which see it as a balance between order (high $S$) and chaos (high $E$).
\textcolor{black}{Since $E$ is a unitless measure and $S$ is measured in bits, the units of $C$ are also bits. C can be interpreted as how much information is reduced ($S$) multiplied by the information production ratio (E).}
Considering the previous simplification of  $I_{in}=1$, we can represent complexity in terms of  $I_{out}$:

\begin{equation}
C=a\cdot I_{out} (1-I_{out}),
\label{eq:Cs}
\end{equation}
where $a$ is a normalizing constant. For the Boolean case, since $I \in [0,1]$, $a=4$ will bound $C \in [0,1]$. An illustration of the simplified versions of $E$, $S$, and $C$ (equations \ref{eq:Es}, \ref{eq:Ss}, and \ref{eq:Cs}) can be seen in Figure \ref{fig:entropy2}.

The Hamming distance $d$ measures the percentage of different bits or symbols in two strings, say $A$ and $B$, of the same length ($|A|=|B|$). For binary strings, it can be calculated with the XOR function ($\oplus$). Its normalization bounds the Hamming distance to the interval $[0,1]$:
\begin{equation}
d(A,B)= \frac{\sum_{i \in [0..|A|]}{a_i \oplus b_i}}{|A|},
\label{eq:d}
\end{equation}
where $a_i$ is the $i^{th}$ bit of string $A$, $b_i$ is the $i^{th}$ bit of string $B$, and $|A|$ represents the length of both strings.
In other words, $d$ measures the fraction of bits that are different between $A$ and $B$. For the Boolean case, $d=0 \iff A=B$ and $d=1 \iff A=\neg B$, while $A$ and $B$ are uncorrelated $\iff d\approx0.5$.

The normalized Hamming distance $d$ can be used to measure homeostasis. If we focus on the information change produced by a process, $d(I_{in},I_{out})$ will measure how much change took place. Homeostasis $H$ can be defined as the opposite:

\begin{equation}
H=1-d(I_{in},I_{out}).
\label{eq:H}
\end{equation}

A high $H$ implies that there is no change, i.e. information is maintained. This occurs naturally in highly ordered systems (with a high $S$), but is also desirable in complex systems (with a high $C$). \textcolor{black}{$d$ and thus $H$ are unitless measures.}

\textcolor{black}{Homeostasis is closely related to sensitivity to initial conditions. A high sensitivity to initial conditions (uncorrelated $H$) is a signature of chaotic dynamics, and can be formally measured with Lyapunov exponents~\citep{Gershenson2004c}. A high $H$ indicates stability.}

\subsection{Multi-scale measures}

To measure emergence, self-organization, complexity, and homeostasis at multiple scales, binary strings can be converted to different bases. If we take two bits, then we will have strings in base 4. If we take three bits, strings will have base 8. Taking $b$ bits will give strings of base $2^b$.
The information of strings can be normalized to $[0,1]$ by dividing Shannon's information by the number $b$ of bits considered:

\begin{equation}
I_b=I/b.
\label{eq:Ims}
\end{equation}

Table \ref{tab:scaled} shows an example of a 32-bit string converted to different bases, each row representing an order of magnitude difference~\citep{Salthe:2012}. Notice that $I_b$ seems to decrease as the base increases. This is a finite size effect, because for large scales most values are not represented in short strings, so their probability is considered as zero. In very long strings, however, this effect is less noticeable, so long random strings have a high entropy for different scales ($I_b\approx 1$), as shown in Table \ref{tab:long}.

\begin{table}[htdp]
\caption{Example of a 32-bit string. Grouping $b$ bits, it can be scaled to different bases.}
\begin{center}
\begin{tabular}{|l|l|l|l|}
\hline
\textbf{$b$}&	\textbf{base} & \textbf{string} & \textbf{$I_b$}\\
\hline
1&	2	&0 0 0 0 1 0 0 0 1 0 1 0 0 0 0 1 1 1 0 0 1 0 0 0 1 1 0 0 1 0 0 0	& 0.89603821\\
2&	4	& 0 0 2 0 2 2 0 1 3 0 2 0 3 0 2 0 & 0.8246987\\
4&	16	&  0  8 10  1 12  8 12  8	&	 0.5389098\\
8&	256	&   8 161 200 200	& 0.1875\\
\hline

\end{tabular}
\end{center}
\label{tab:scaled}
\end{table}%

\begin{table}[htdp]
\caption{Example of $I_b$ for a long pseudorandom string at different scales.}
\begin{center}
\begin{tabular}{|l|l|l|l|}
\hline
\textbf{$b$}&	\textbf{base} & \textbf{length} & \textbf{$I_b$}\\
\hline
1&	2	& $2^{20}$	& 0.9999998\\
2&	4	& $2^{19}$	& 0.9999997\\
4&	16	& $2^{18}$	& 0.9999956\\
8&	256	& $2^{17}$ 	& 0.9998395\\
\hline

\end{tabular}
\end{center}
\label{tab:long}
\end{table}%

$E$, $S$, $C$, and $H$ for different scales can be obtained converting binary strings to higher bases \textcolor{black}{and normalizing their information with Eq. \ref{eq:Ims}}. The behavior is similar for all except $H$, since the Hamming distance measures the percentage of different symbols between strings. For binary strings ($b=1$), uncorrelated strings have a Hamming distance $d \approx 0.5$, since the probability of having the same symbol in each position is 0.5. For base 4 strings  ($b=2$), this will be halved to $d \approx 0.25$, because the probability of having the same symbol is each position is also halved to 0.25. Thus, the lowest expected $H$ (uncorrelated states) will decrease with increasing $b$ in the form of $\frac{1}{2b}$.

Since $I$ can change drastically depending on the scale at which it is observed (e.g. a string $10101010$ has $I_1=1$ but $I_2=0$, since the string in base 4 becomes $2222$), all the proposed measures can also have drastic changes with a change of scale.

\section{Experiments}
\label{sec:experiments}

In this section we present computational experiments  designed to exemplify the measures proposed above, as well as studying their changes with scale.

\subsection{Random Boolean Networks and Elementary Cellular Automata}

Random Boolean networks (RBNs) were first proposed by Stuart Kauffman as models of genetic regulatory networks~\citep{Kauffman1969,Kauffman1993} and have been extensively studied~\citep{Wuensche1998,Gershenson2002e,AldanaEtAl2003,Gershenson2004c,Drossel:2009,Gershenson:2010}.

A RBN consists of $N$ Boolean nodes, i.e. they can take values of 0 or 1. The state of each node is determined by $K$ other nodes (on average). The \emph{structure} of the RBN is determined by the network of interacting nodes. The \emph{function} of the RBN is given by lookup tables that determine the future state of each node depending on the current state of its inputs. The structure and the function of RBNs are initially generated randomly and remain fixed afterwards.

Classic RBNs are updated synchronously~\citep{Gershenson2004b} and thus their dynamics are deterministic. Since their state space is finite ($2^N$ states), sooner or later a state will be repeated. When this occurs, the RBN has reached an attractor. These can be point attractors (a single state) or cycle attractors of different lengths.

Depending on different structural and functional properties, RBN dynamics can be ordered, chaotic, or critical~\citep{Gershenson:2010}. Ordered dynamics are characterized by few \textcolor{black}{nodes changing} and high robustness. Chaotic dynamics are characterized by \textcolor{black}{most nodes changing} and high fragility. Critical dynamics balance the variability of the chaotic regime with the robustness of the ordered regime. Critical dynamics are related to a high complexity.

Elementary cellular automata (ECA) have also been studied extensively~\citep{Wolfram:1984,WuenscheLesser1992,Wolfram:2002}. They can be seen as particular cases of RBNs~\citep{Wuensche1998,Gershenson2002e}, where all nodes have the same function (rule) and the structure is regular, i.e. each node has $K=3$, inputs: themselves and their closest neighbors. There are 256 possible ``rules" (the possible combinations of Boolean functions for 3 inputs are $2^{2^3}=2^8=256$), although only 88 equivalence classes~\citep{Li:1990,WuenscheLesser1992}. There have been several classifications of ECA dynamics, the most popular being Wolfram's~\citeyearpar{Wolfram:1984}. Roughly, class I and II rules generate ordered dynamics. Rules from class I tend to point attractors from all initial states, while class II rules tend to cycle attractors. Class III rules have chaotic dynamics. Rules from class IV generate critical dynamics, and some have been proven to be capable of universal computation~\citep{Cook2004}. 

There are several methods to measure criticality in RBNs, including Derrida's annealed approximation~\citep{DerridaPomeau1986}, Lyapunov exponents~\citep{LuqueSole2000}, and Hamming distances~\citep{Gershenson2004a}. 

The study of complexity and related properties of discrete dynamical systems with information theory is becoming a very active area.
For example, Fisher information was used recently to measure criticality in RBNs~\citep{wang2011fisher}. Zenil~\citeyearpar{Zenil:2010} proposed a compression-based method to detect phase transitions (which are related to complexity).
Novel measures of complexity of RBNs have been proposed by \textcolor{black}{Gong and Socolar}~\citeyearpar{Gong:2012}.
Mart\'inez et al.~\citeyearpar{Martinez:2012} study CA with memory, being able to extract critical dynamics from chaotic rules.

\subsection{Experimental setup}

Several experiments were performed to explore the measures proposed in Section \ref{sec:proposal}. To measure emergence $E$, self-organization $S$, and complexity $C$, the time series of single nodes were evaluated in equations  \ref{eq:Es}, \ref{eq:Ss}, and \ref{eq:Cs} and then averaged to obtain the results for the network. \textcolor{black}{To measure homeostasis $H$, $d$ was measured between the last two states.}
 1000 RBNs ran from a random initial state for 1000 steps and then 1000 ``additional" steps were used to generate results\textcolor{black}{: the additional states of each node were used to calculate $I_{out}$ for each node. These were averaged to calculate $I_{out}$ for the network}. For ECA, fifty instances of some rules of the four Wolfram classes were evaluated, considering 256 nodes, $2^{12}$ initial steps and $2^{12}$ additional steps were used to generate results.

The simulations were programmed in R~\citep{R}, using packages BoolNet~\citep{Mussel:2010},
CellularAutomaton~\citep{Hugues:2012}, and entropy~\citep{Hausser:2012}. 

\subsection{Results}

\begin{figure}[htbp]
\begin{center}
  \includegraphics[width=0.95\textwidth]{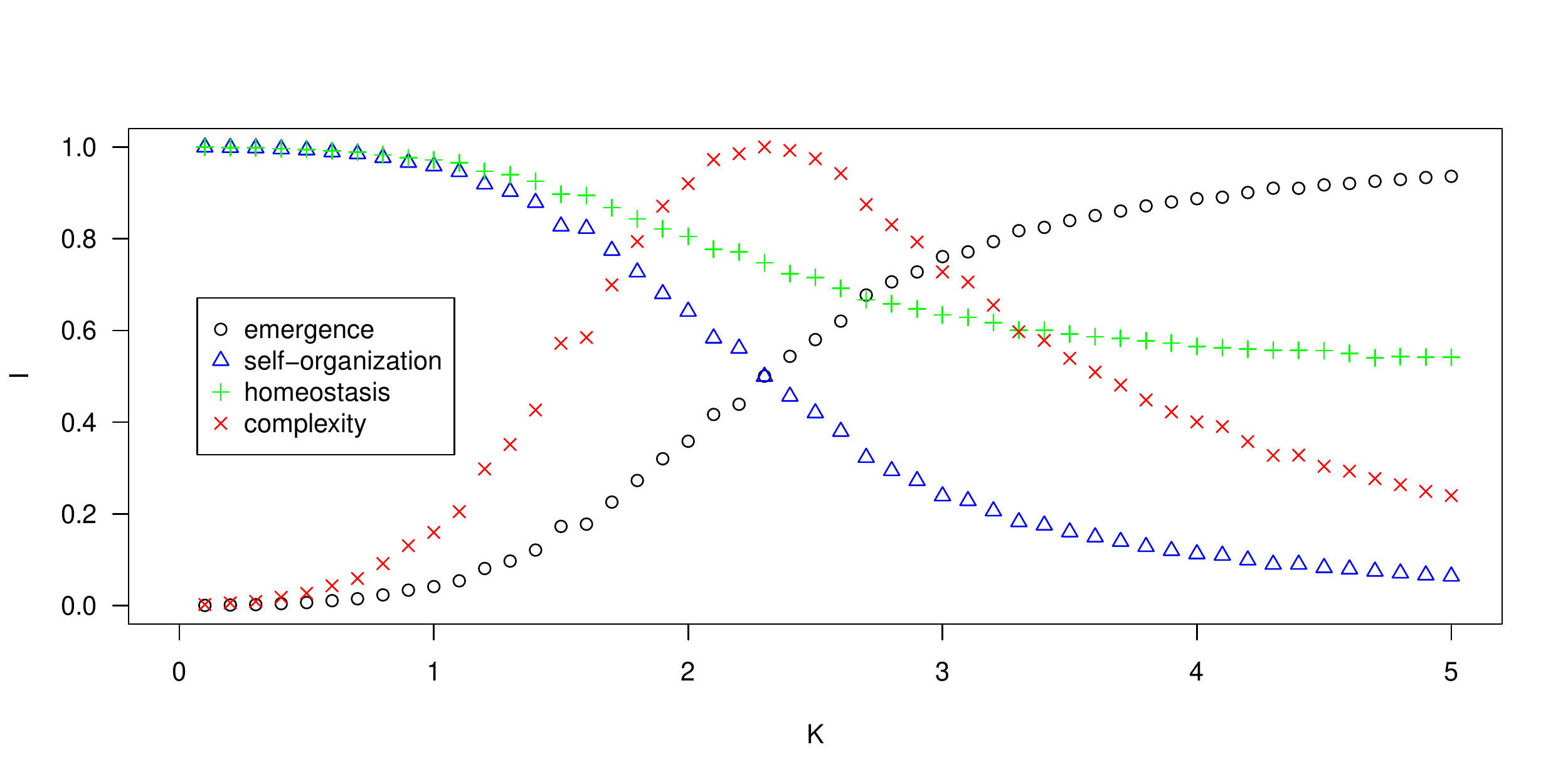}\\
\caption{Averages for 1000 RBNs, $N=100$ and varying $K$.}
\label{InfoRBN-pl}
\end{center}
\end{figure}

\begin{figure}[htbp]
\begin{center}
  \includegraphics[width=0.95\textwidth]{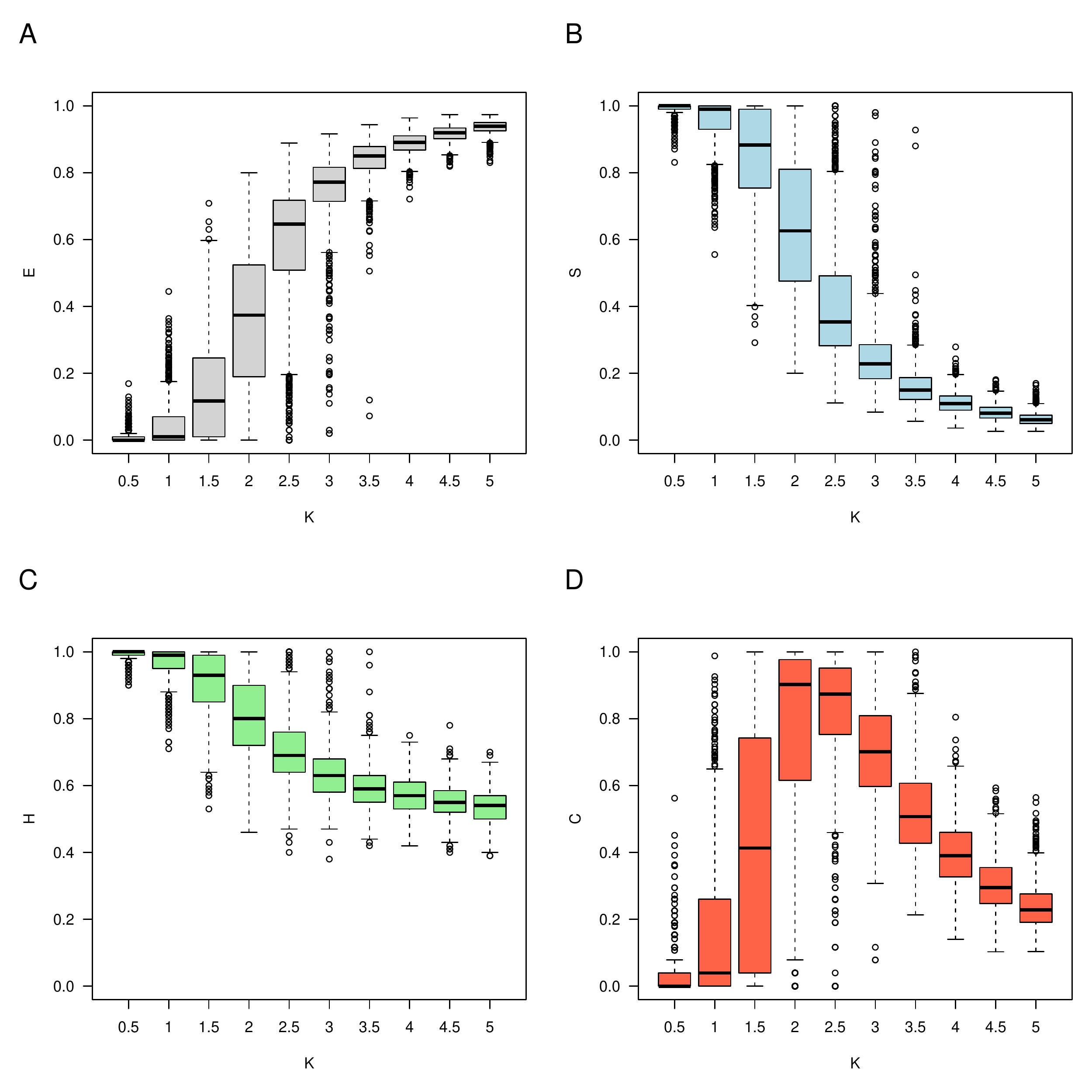}\\
\caption{Boxplots for 1000 RBNs, $N=100$ and varying $K$. A. Emergence. B. Self-organization. C. Homeostasis. D. Complexity.}
\label{InfoRBN-bp}
\end{center}
\end{figure}

Figures \ref{InfoRBN-pl} and \ref{InfoRBN-bp} show experimental results for RBNs. It can be seen that for low connectivity $K$, emergence $E$ and complexity $C$ are low, while self-organization $S$ and homeostasis $H$ are high. These are features of ordered dynamics, i.e. few changes. For high $K$, almost the opposite is observed: $E$ is high and $S$ and $H$ are low. However, $C$ is also low. These are characteristics of chaotic dynamics, i.e. high variability. For medium connectivities ($2<K<3$), there is a balance between $E$ and $S$, giving high values of $C$. This is consistent with critical dynamics observed in RBNs\textcolor{black}{~\citep{Gershenson2004c}, where criticality is found theoretically at $K=2$ and for finite systems at connectivities $2<K<3$}. The values of $H$ are also a balance between no change ($H=1$) and no correlation ($H \approx 0.5$).

Experiments were also performed for higher scales of observation\textcolor{black}{, considering successive states of nodes as single higher scale states}. These are not shown, since there was not a noticeable change, except for a slight shift of the maximum $C$ value, which increases moderately with the scale and a slight decrease of $E$ for higher scales due to the limited size of evaluated strings (1000 bits per node). Low $H$ values (non-correlation) decrease with scale, as explained above.

\begin{figure}
     \centering
     \subfigure[]{
          \label{fig:ECAA}
          \includegraphics[width=.22\textwidth]{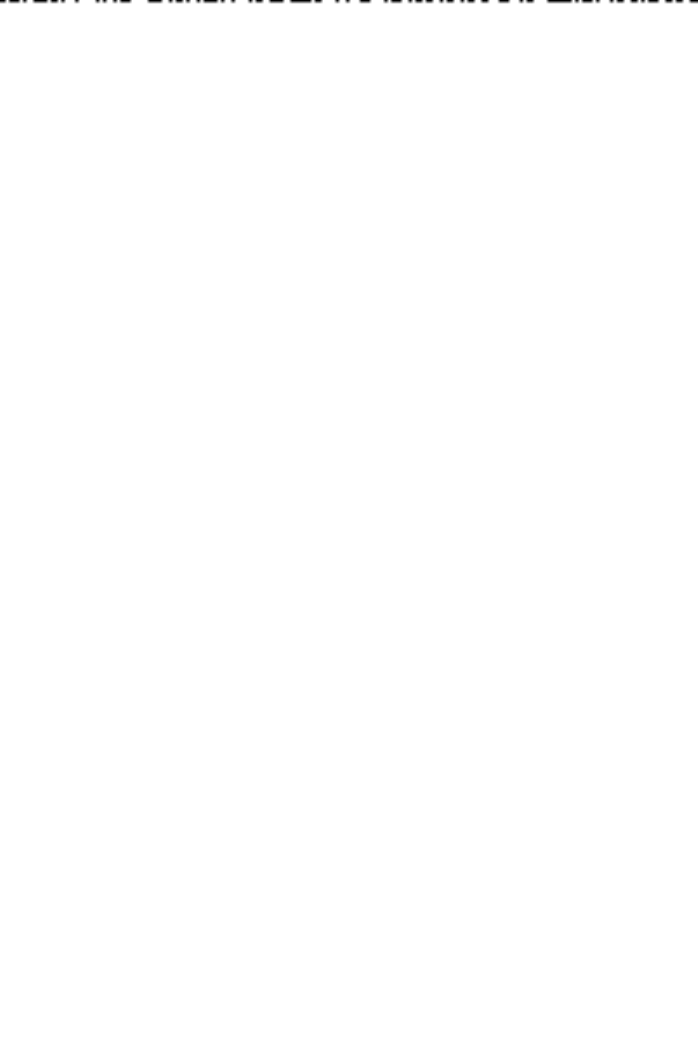}
	}
     \subfigure[]{
          \label{fig:ECAB}
          \includegraphics[width=.22\textwidth]{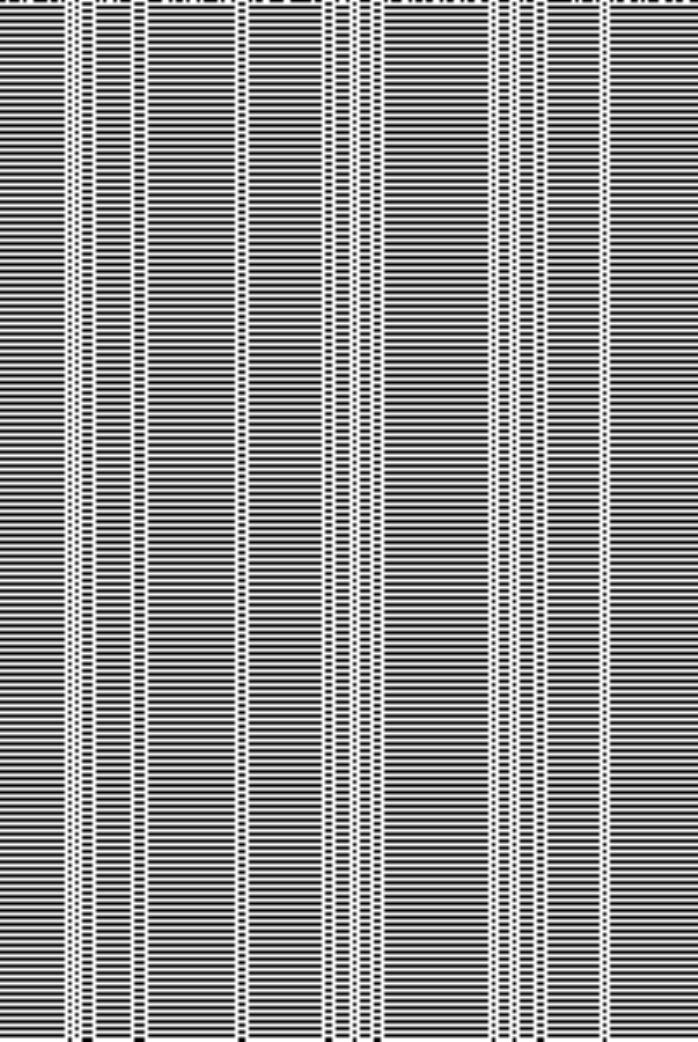}
	}
     \subfigure[]{
          \label{fig:ECAC}
          \includegraphics[width=.22\textwidth]{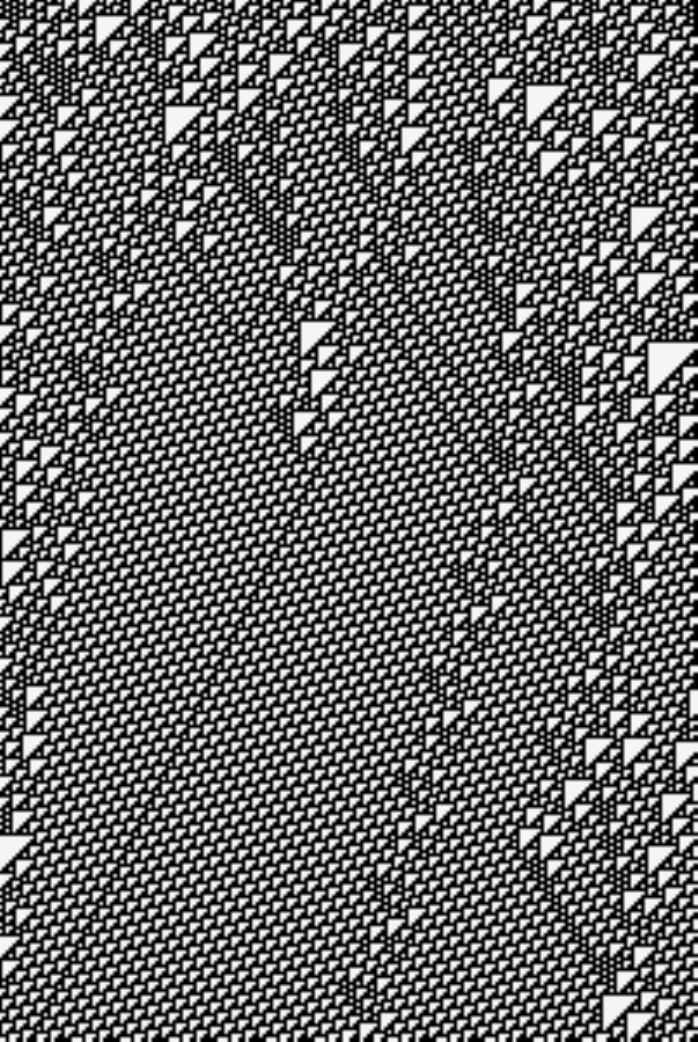}
	}
     \subfigure[]{
          \label{fig:ECAD}
          \includegraphics[width=.22\textwidth]{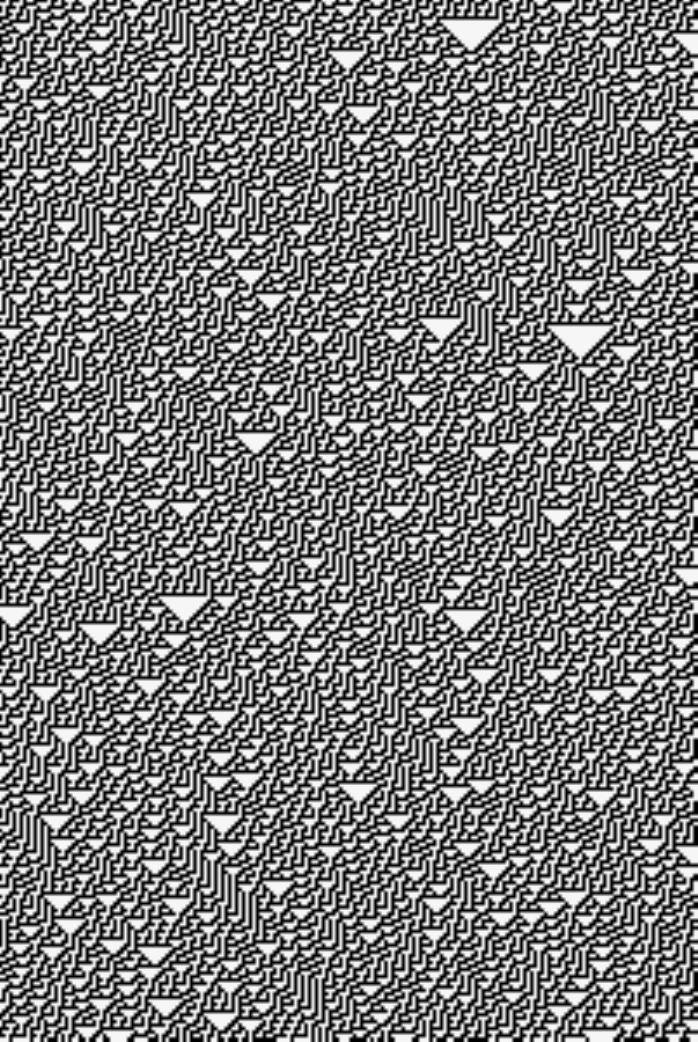}
	}

     \caption{\textcolor{black}{Examples of ECA with a random initial state (top row): (A) rule 0 from class I, (B) rule 1 from class II, (C) rule 110 from class IV, and (D) rule 30 from class III.
Images obtained from \url{http://wolframalpha.com}. }   }
     \label{fig:rECA}
\end{figure}

\begin{figure}[htbp]
\begin{center}
  \includegraphics[width=0.85\textwidth]{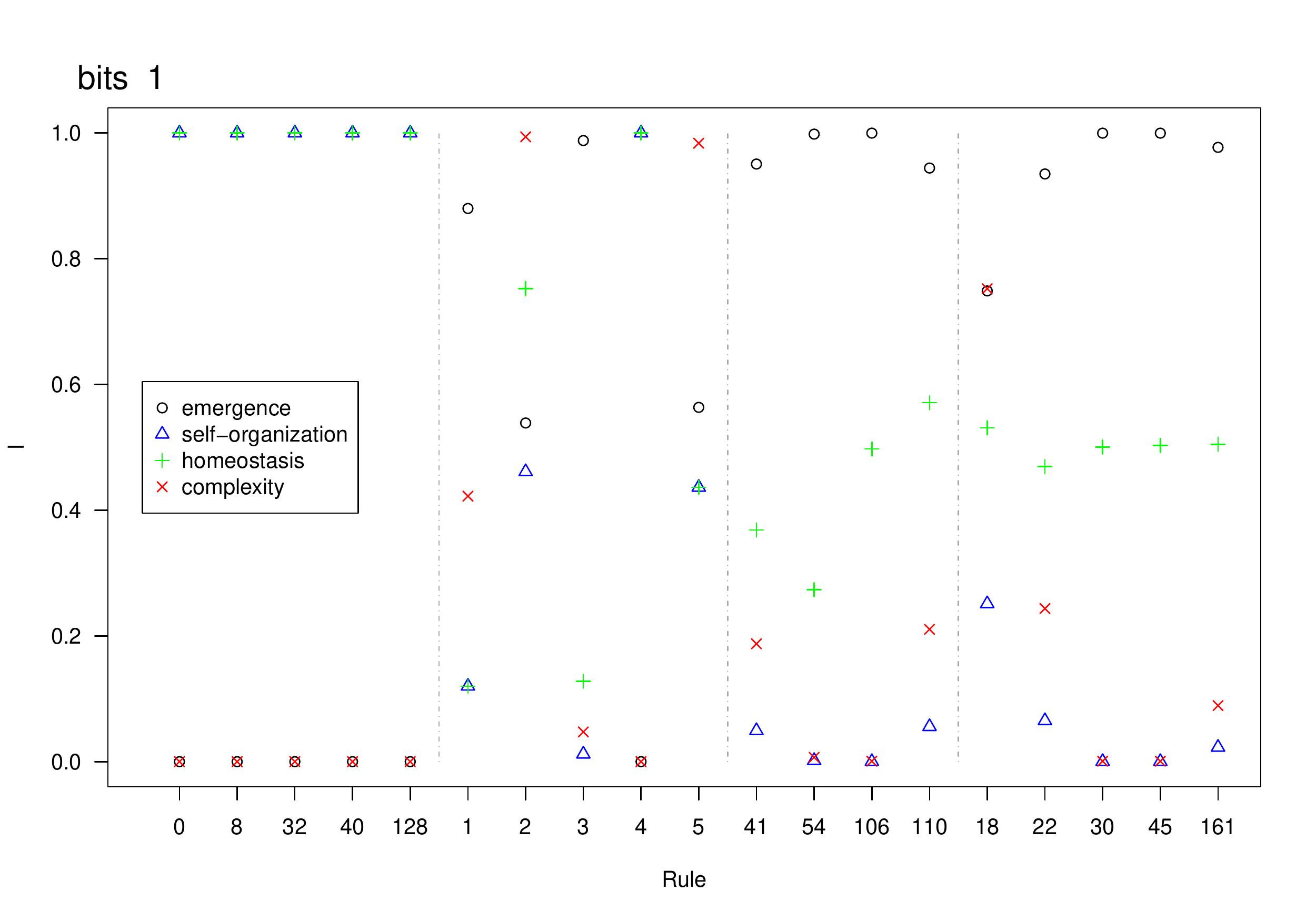}\\
\caption{Averages of 50 ECA for 19 rules, $N=256$, $b=1$.}
\label{InfoCAb1}
\end{center}
\end{figure}

\begin{figure}[htbp]
\begin{center}
  \includegraphics[width=0.85\textwidth]{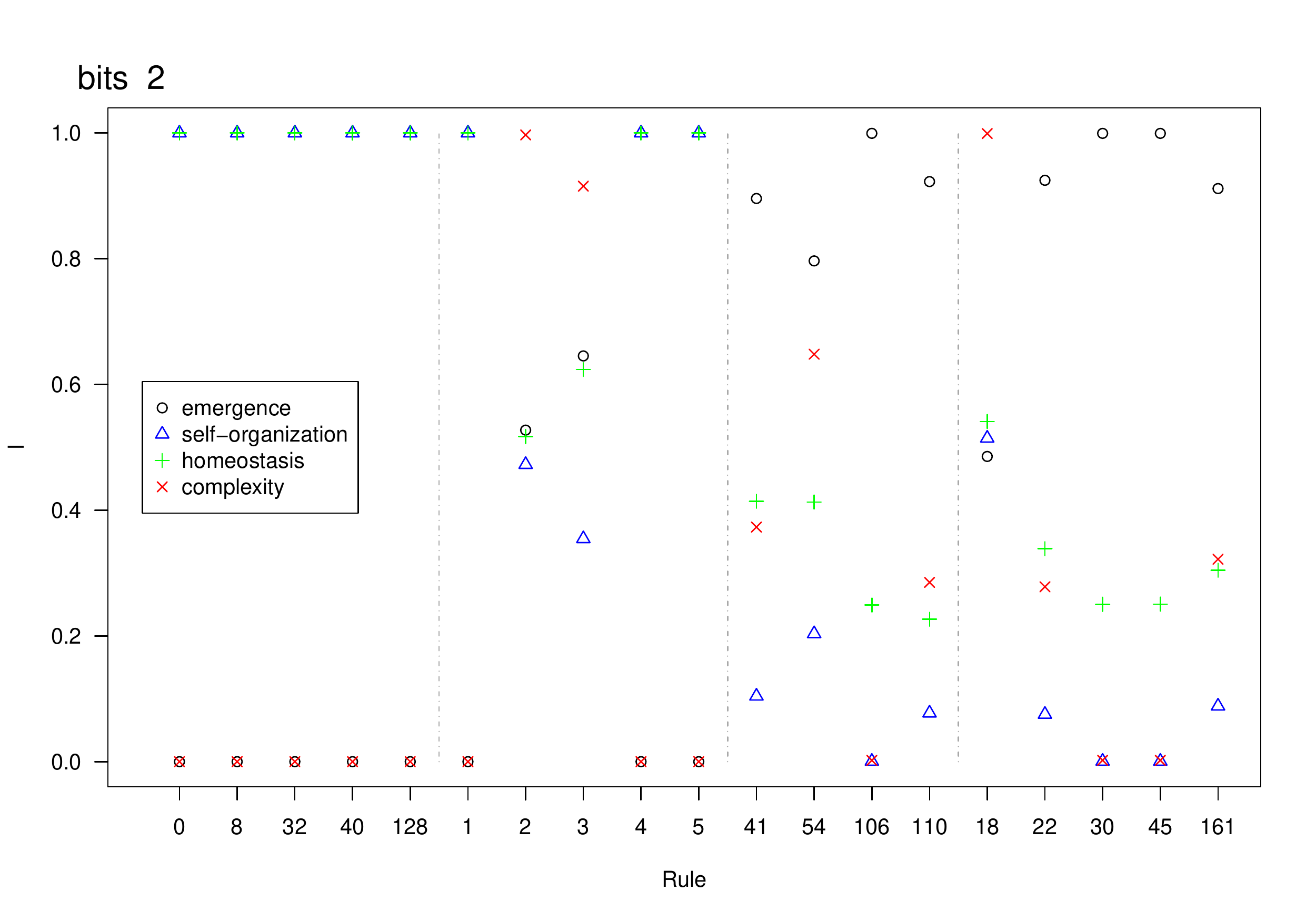}\\
\caption{Averages of 50 ECA for 19 rules, $N=256$, $b=2$.}
\label{InfoCAb2}
\end{center}
\end{figure}

\begin{figure}[htbp]
\begin{center}
  \includegraphics[width=0.85\textwidth]{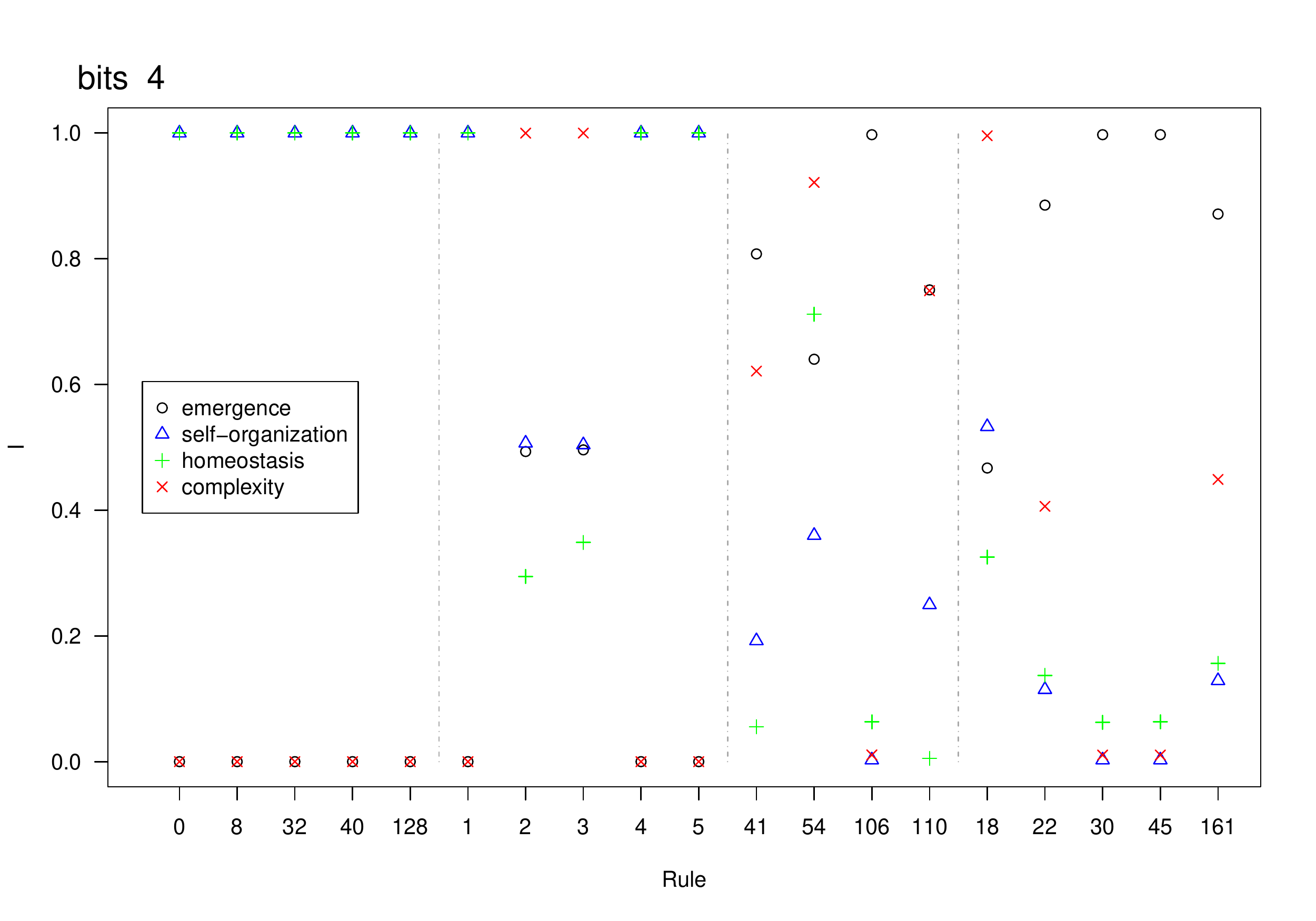}\\
\caption{Averages of 50 ECA for 19 rules, $N=256$, $b=4$.}
\label{InfoCAb4}
\end{center}
\end{figure}

\begin{figure}[htbp]
\begin{center}
  \includegraphics[width=0.85\textwidth]{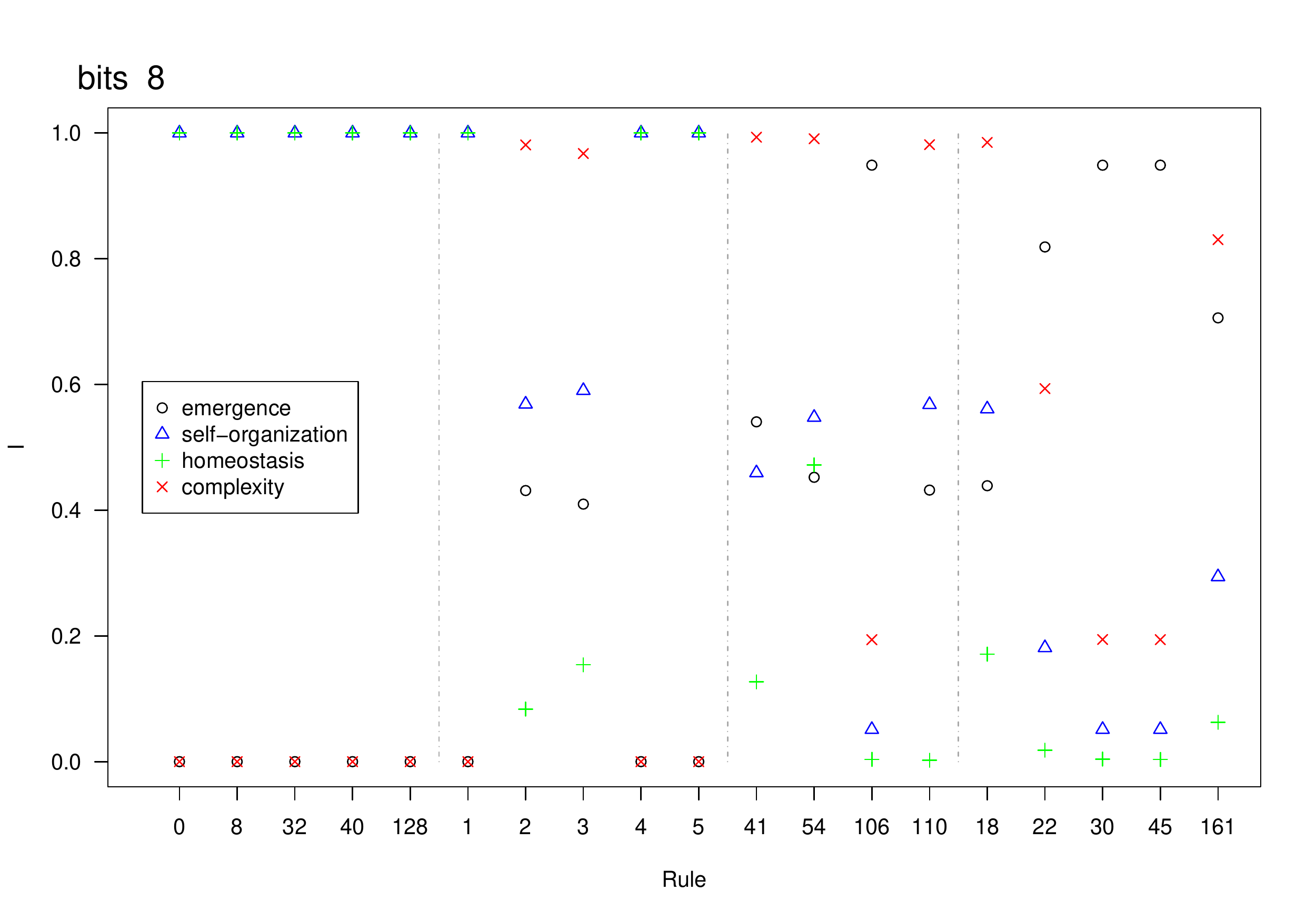}\\
\caption{Averages of 50 ECA for 19 rules, $N=256$, $b=8$.}
\label{InfoCAb8}
\end{center}
\end{figure}

19 ECA rules were evaluated corresponding to different classes, each belonging to a different equivalence class: Rules 0, 8, 32, 40, and 128 from class I, rules 1, 2, 3, 4, and 5 from class II, rules 18, 22, 30, 45, and 161 from class III, and rules 41, 54, 106, and 110 from class IV (there are only four equivalence classes of type IV). Results for different scales ($I_1$, $I_2$, $I_4$, and $I_8$) are shown in Figures \ref{InfoCAb1}--\ref{InfoCAb8}. \textcolor{black}{Example patterns from one ECA of each class are shown in Figure \ref{fig:rECA}.}

It can be seen that rules from class one have a $E=C=0$ and $S=H=1$ for all scales. This is because there is no change in any node after the ECA reaches a stable state (which usually takes only one time step).

Some rules from class II behave similarly to class I, e.g. rule 4. Other rules---such as rules 1 and 5---have a relatively high $E>0.5$ at $b=1$. This is because the majority of cells are oscillating between zero and one at every time step. However, as the scale is increased ($b \geq 2$), these patterns become regular and also behave as class I rules ($E=C=0$ and $S=H=1$). Other rules---such as rules 2 and 3--- have regular diagonal patterns. Since we are measuring information change per node, they appear to have a high $C$. This is because in practice the ECA is transmitting information globally, although this information is not transformed. If information was measured diagonally (in the direction of the dynamics), these rules would have similar properties as those of class I rules.

Rules from class III tend to have a high $E$. Rule 18 is particular, since it has a larger proportion of zeros than ones, being accumulated in triangles. Even when the occurrence and size of triangles is difficult to predict, this reduces the $E$ and increases the $S$, $C$, and $H$ of rule 18 compared to other class III rules. This is noticeable even more at higher scales, since other class III rules have a more balanced percentage of zeros and ones. Rule 161 is in between, having a slightly larger percentage of ones than zeroes. 

Rules from class IV behave similarly to class III rules for $b=1$, with a high $E$, low $S$ and $C$. Rule 54 is an exception for $H$, where the alternating pattern of the ether of period 4\footnote{The regular pattern of complex ECA is known as ether. Persistent structures are known as gliders, which navigate the ether without producing changes in the patterns of either.} generates an anticorrelated $H$. The regular ether of rule 54 and also the relative regularity of its gliders decrease $E$ for higher scales, with a very high $H$ for $b \geq 4$. Rule 106 has diagonally interacting patterns, which lead to a large $E$ at all scales, for reasons similar to those of rules 2 and 3. Rule 41 decreases its $E$ and increases its $S$ and $C$ with scale, and maintains a non-correlated $H$. This reveals more regular patterns at higher scales, similar to rule 18. Rule 110 behaves almost the same as rule 41: $E$ is reduced and $S$ and $C$ increase for higher scales. The difference lies in $H$. For rule 110, $H$ is anticorrelated for $b \geq 4$. This again can be explained with the ether, which has a period 7 for rule 110.   

\section{Discussion}
\label{sec:discussion}

\begin{figure}
     \centering
     \subfigure[]{
          \label{fig:ECAA}
          \includegraphics[width=.45\textwidth]{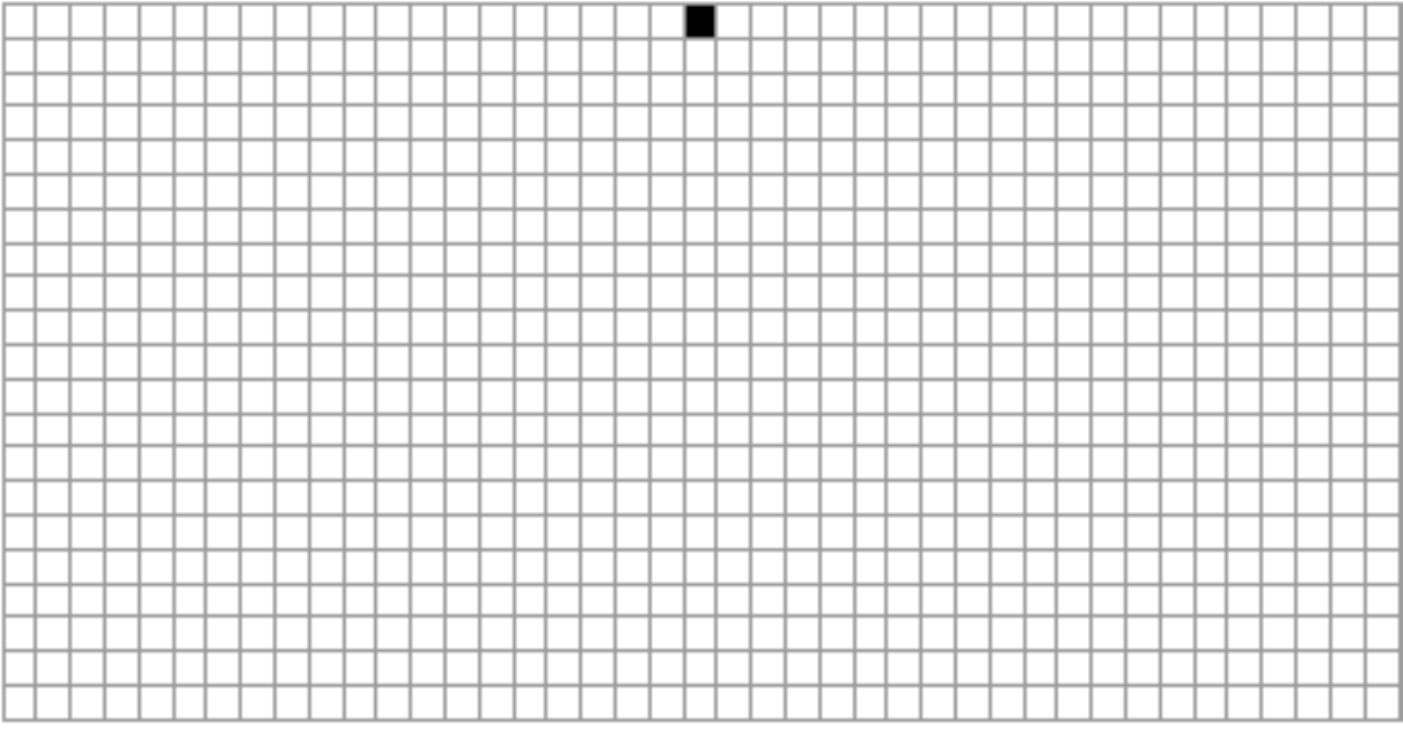}
	}
     \subfigure[]{
          \label{fig:ECAB}
          \includegraphics[width=.45\textwidth]{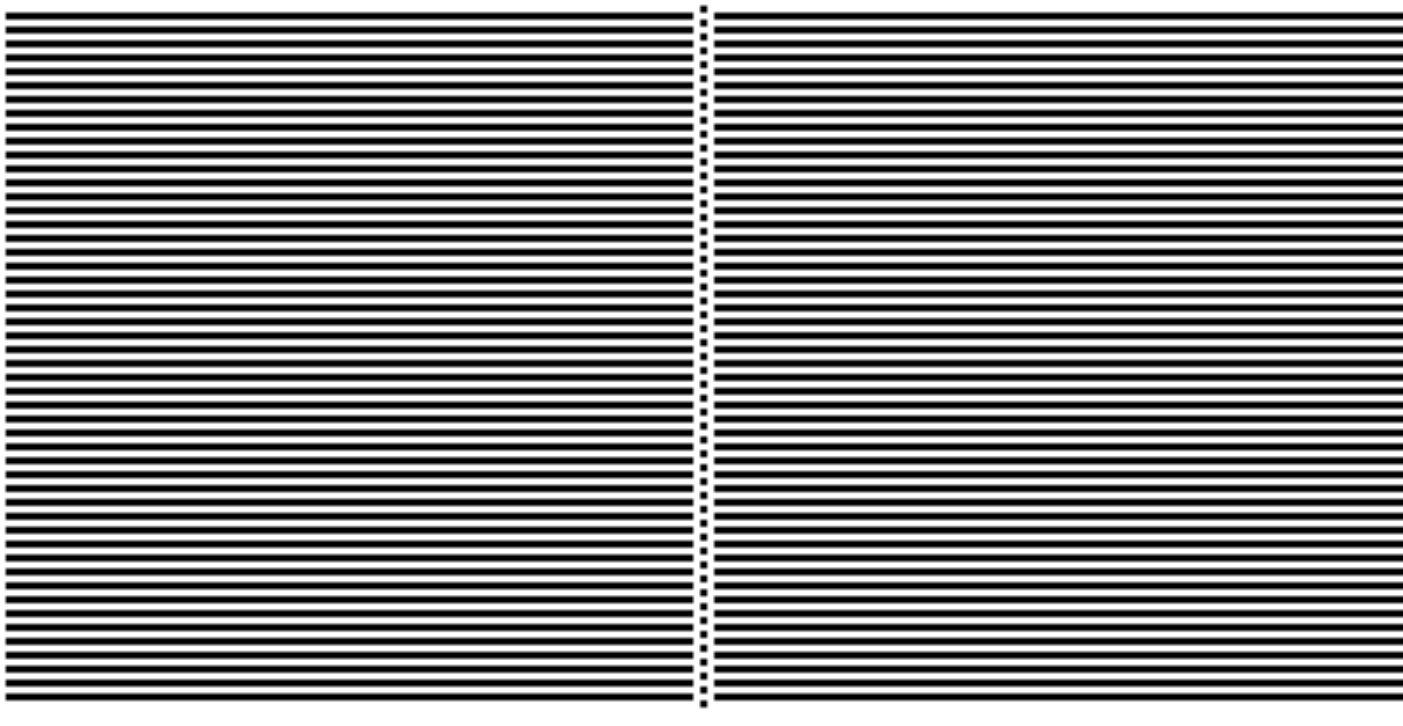}
	}
\\
     \subfigure[]{
          \label{fig:ECAC}
          \includegraphics[width=.45\textwidth]{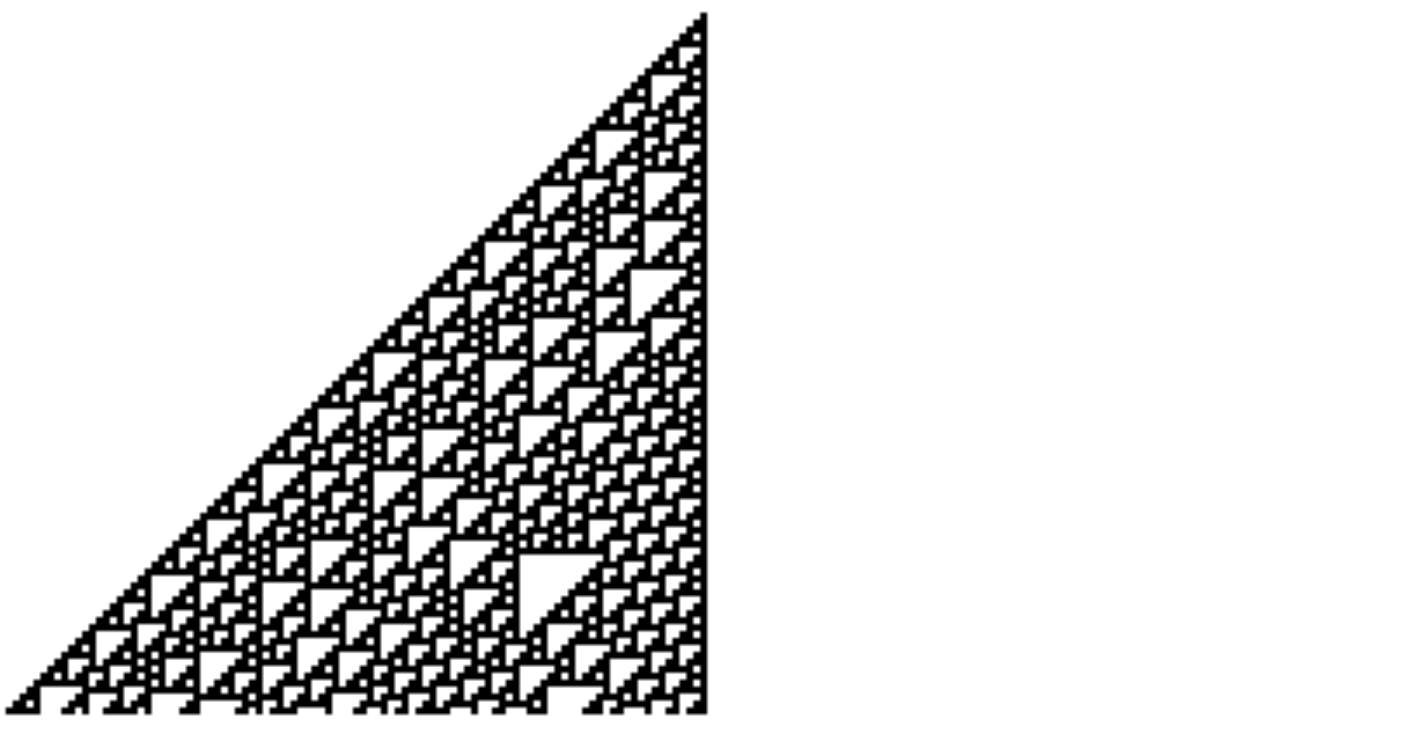}
	}
     \subfigure[]{
          \label{fig:ECAD}
          \includegraphics[width=.45\textwidth]{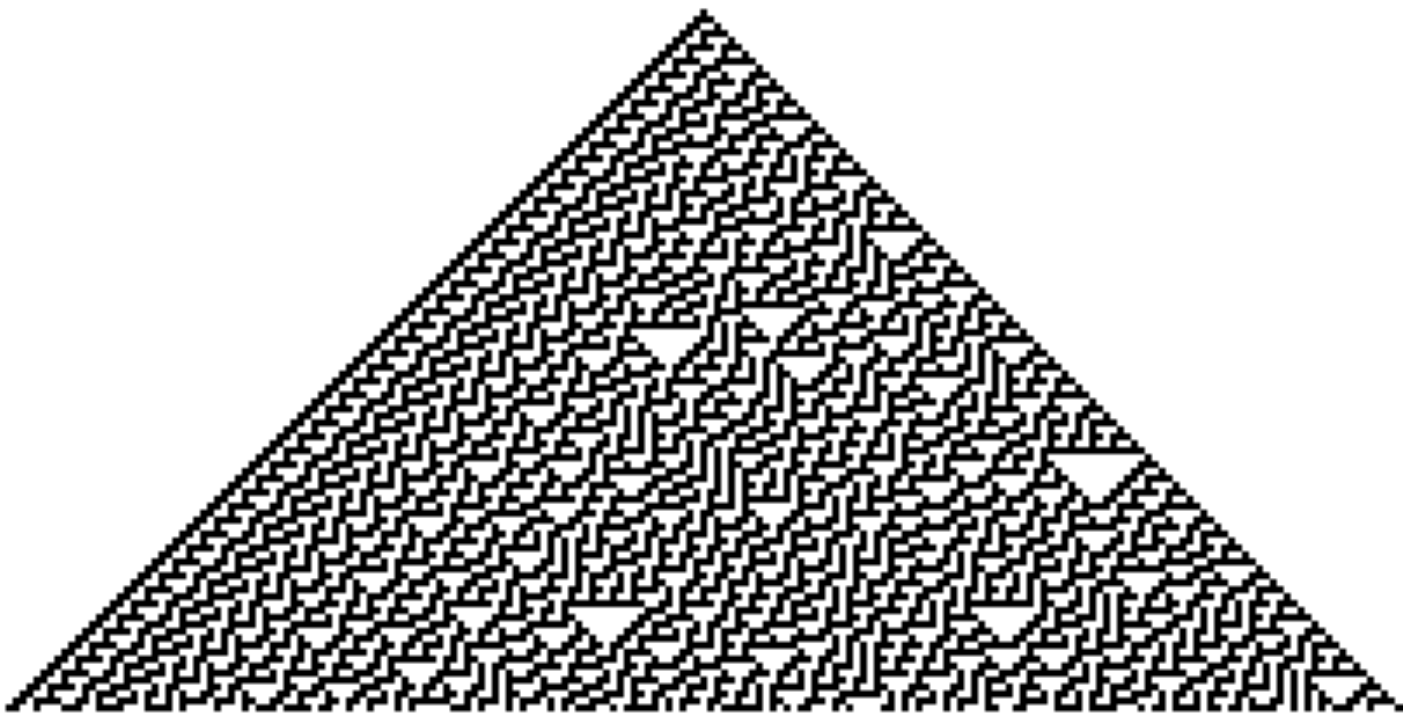}
	}

     \caption{\textcolor{black}{Examples of ECA with a single initial cell: (A) rule 0 from class I, (B) rule 1 from class II, (C) rule 110 from class IV, and (D) rule 30 from class III. 100 steps are shown except for (A), only 20 steps.
Images obtained from \url{http://wolframalpha.com}.   } }
     \label{fig:ECA}
\end{figure}

\textcolor{black}{The proposed measure of emergence considers random processes as being those with the highest $E$, because there is more \emph{information} emerging. This might not be clearly illustrated by the experiments which use the simplified $I_{in}=1$, making the emergence (ratio between produced information over introduced information) to lie in the interval $[0,1]$.
ECA with an initial condition of a single ``on" cell (minimum $I_{in}>0$) can be used to visually exemplify $E$ more concretely, as shown in Figure \ref{fig:ECA}. Rule 0 is class I, and no matter how much information is fed into the ECA, this is lost after one time step. Minimum emergence: anything comes in, nothing comes out. Rule 1 from class II produces a lot of information for $b=1$, where a stripped pattern emerges. However, for $b \geq 2$ this pattern disappears, so the emergence is also minimal at those scales. Rule 110 from class IV produces complex patterns and structures, with gliders that interact on a regular (ether). We can say that these patterns are emerging, as indicated by the increase of information produced by the system, compared to that introduced. Rule 30 from class III produces even more patterns, since its pseudorandom behavior delivers maximum information produced. Maximum emergence: get the most information putting the least in. Measuring emergence at multiple scales at once with an ``emergence profile" can give more insight about the dynamics of a system.}

\begin{figure}[htbp]
\begin{center}
  \includegraphics[width=0.95\textwidth]{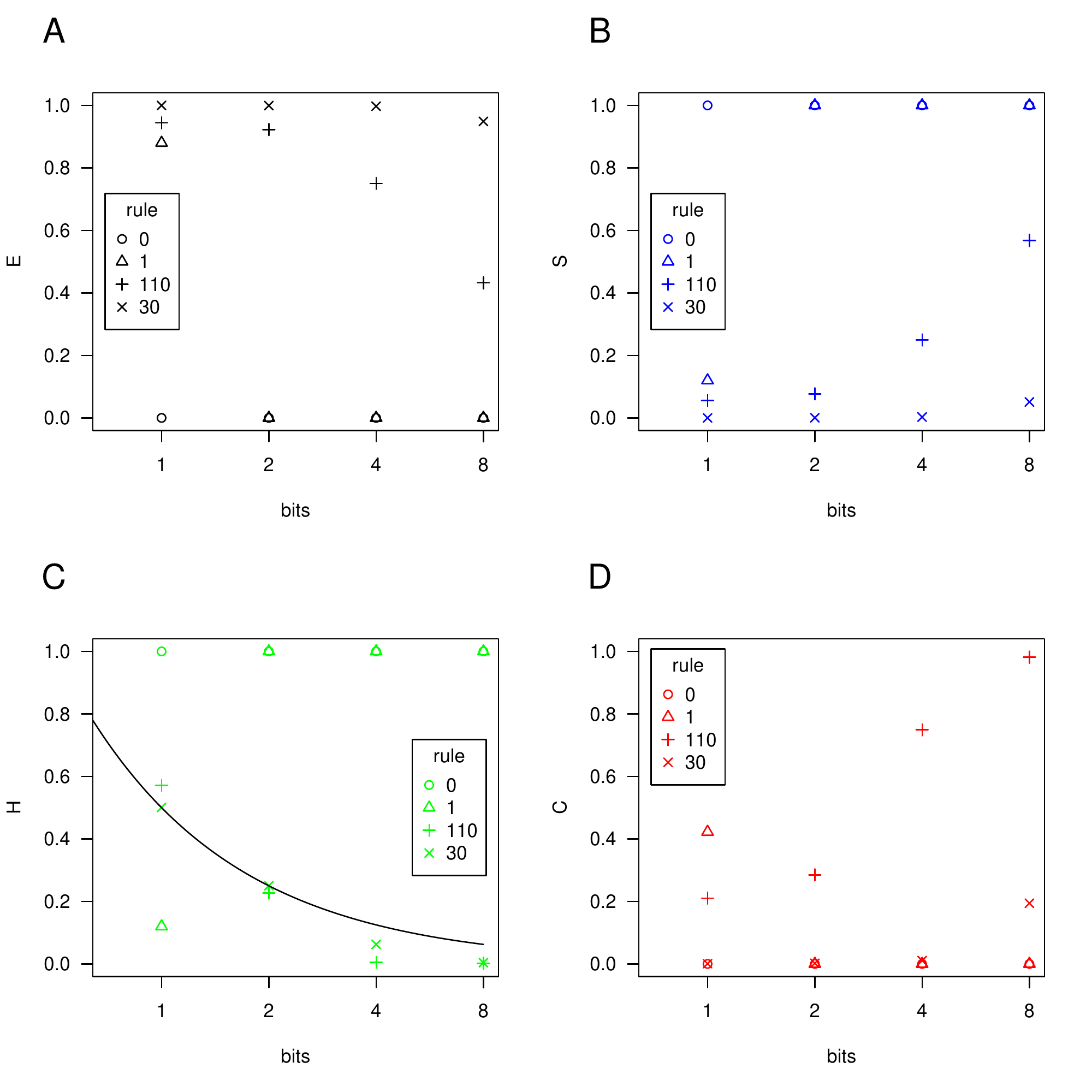}\\
\caption{\textcolor{black}{Multiscale profiles of rules 0, 1, 110 and 30. A. Emergence. B. Self-organization. C. Homeostasis (line shows $\frac{1}{2b}$). D. Complexity.}}
\label{InfoCA-ms}
\end{center}
\end{figure}

\textcolor{black}{A multiscale profile for different measures and four ECA rules is shown in Figure \ref{InfoCA-ms}. Rule 0 does not change with scale. Rule 1 has high $E$, low $S$ and $H$ (anticorrelated), and medium $C$ for $b=1$, but is similar to rule 0 for higher scales. Rule 110 decreases its $E$ and $H$ and increases its $S$ and $C$ with scale. Rule 30 has a high $E$ and a low $S$, $H$, and $C$ for all scales.}

\textcolor{black}{Self-organization is defined as a reduction of Shannon information, which is also considered as entropy or disorder. Minimum self-organization occurs when a highly ordered pattern (minimum $I_{in}$) is converted into a highly disordered pattern (maximum $I_{out}$). Chaotic systems such as rule 30 exhibit minimal self-organization. Maximum self-organization occurs when a highly disordered pattern is converted into a highly ordered pattern. Systems which tend quickly to a steady state, such as class I ECA, exhibit maximum self-organization. 
Shalizi et al.~\citeyearpar{Shalizi:2004} define self-organization as an internal increase of statistical complexity. This measure would give systems with a high $C$ maximum self-organization, while we consider a maximum $S$ for static systems. We believe that a frozen state is more \emph{organized} than a complex pattern, in line with thermodynamics and information theory. However, a system with a minimum $S$ most probably is not \emph{organizing}. We believe that the measure of Shalizi et al. is more appropriate for studying the dynamics of self-organizing systems~\citep{GershensonHeylighen2003a,GershensonDCSOS}, while $S$ is intended as a general measure of self-organization. 
}

From our simplified definitions, self-organization can be seen as the opposite of emergence: self-organization is high for ordered dynamics, emergence is high for chaotic dynamics (at all scales). \textcolor{black}{Having self-organization and emergence as opposites might seem counterintuitive, since there are several systems which seem to have a high emergence \emph{and} self-organization. Precisely these systems seem to be those most interesting to science. We agree, and these systems are those with a highest $C$, but if one goes to the extremes of $E$ or $S$, one finds extreme disorder or extreme order, respectively.}

Information (and emergence) can be seen as the balance of zeros and ones ($P(0)=1-P(1)$; $max(I) \iff P(0)=P(1)=0.5$). Complexity can be seen as the balance of emergence and self-organization (\textcolor{black}{if $I_{in}=0$}, $S=1-E$; $max(C) \iff S=E=0.5$). This has already been noted as complexity being maximal when there is a tradeoff between persistence of information (memory) and its variability (exploration, computation)~\citep{Langton1990,Kauffman1993}. It has also been argued that there is a tendency for systems to self-organize or evolve to states of increased balance, i.e. high $C$ (at multiple scales)~\citep{BTW1987,Kauffman2000,Gershenson:2007}. This is in agreement with recent work by Escalona-Mor\'an et al.~\citeyearpar{Escalona-Moran:2012}.

\textcolor{black}{Even when the mathematical formulation is different, $C$ is strongly related with Crutchfield et al.'s statistical complexity~\citep{Crutchfield:1989}, which measures how much computation is required to produce information. Ordered and random strings require computation, while complex strings require more computation. This statistical complexity can also be defined as the minimal information required to predict the future of a process~\citep{Shalizi:2004}. For our $C$, low $I$ implies strings which are easy to predict, while high $I$ implies strings where there is not much to be predicted. Both cases give a low $C$ and statistical complexity. High $C$ occurs when $I$ is medium, which corresponds with more computation required to produce or predict such a pattern. Even when $C$ and statistical complexity exhibit similar behaviors, $C$ is much simpler to calculate and explain.
}

Complexity has been confused with chaos, i.e. a high entropy (Shannon information)~\citep{ProkopenkoEtAl2007}. For example, \textcolor{black}{Bar-Yam's} complexity profile measures complexity as the amount of information required to describe a system at different scales~\citep{BarYam2004}. From our perspective, this could be better named ``information profile" or ``emergence profile". 

Inspired by Bar-Yam's complexity (information) profile, the $\sigma$ profile was proposed to measure organization at different scales~\citep{Gershenson:2010a}. Following the approach of L\'opez-Ruiz et al.~\citeyearpar{LopezRuiz:1995} which we used in this paper, these two profiles could be combined to develop a new complexity profile. This would illustrate general properties of complexity at different scales.

In the ECA experiments, it could be seen that rules mainly fall in two categories: either they compute something ($E>0$) or they don't ($E=0$). This is related to Wolfram’s principle of computational equivalence~\citep{Wolfram:2002,Zenil:2010}, which conjectures that dynamical systems are either capable of universal computation (complex) or they are not (simple). From ECA, only rule 110 has been proven to be capable of universal computation~\citep{Cook2004}, but other rules possibly are also capable. Not only from class IV, but also from class III, although the variability makes it difficult to store information. Still, there are techniques to achieve complex dynamics of chaotic systems, e.g. with memory~\citep{Martinez:2012} or with regular external signals~\citep{LuqueSole1997b,LuqueSole1998} (chaos control). Class II might be considered too regular for universal computation, but adding noise\footnote{Gabriel Barello, personal communication.} or a complex signal might enable them also to perform universal computation. Class I is too static, it would require most of the computation to come from the outside of the system. In general, one can employ different techniques to drive a dynamical system towards a high complexity~\citep{Gershenson:2010}.

In most cases, homeostasis is related to self-organization, since a high $S$ indicates low variability, which is also a characteristic of a high $H$, while low $S$ in most cases is accompanied by an uncorrelated $H \approx \frac{1}{2b}$. \textcolor{black}{This is also a signature of chaotic dynamics (high $E$).}
 Nevertheless, some CA with a low $S$ have a correlated $H > \frac{1}{2b}$ or anticorrelated $H < \frac{1}{2b}$, which is a characteristic of complex structures interacting on a regular background (ether). In this respect, different ECA rules can have high $C$ at a particular scale because $S \approx 0.5$. However, rules with a $H$ deviating from $\frac{1}{2b}$ show a signature of complex structures on an ether. The ether eases the computation, so universality \textcolor{black}{can} be explored more easily. This deviation of \textcolor{black}{$H \approx \frac{1}{2b}$} could also be used to explore the immense space of possible dynamical systems. Using principal component analysis, we also found that for RBNs the standard deviation of $H$ is correlated with $C$: RBNs in the ordered regime have consistently high $H$, RBNs in the chaotic regime have consistently uncorrelated $H$, while RBNs in the critical regime have a variable $H$. This is also related to Wuensche's $Z$ parameter, which measures the variance of input entropy over time~\citep{Wuensche1999}. 

\textcolor{black}{Multiscale profiles of the measures give more insight than using them at particular scales, as shown in Figure \ref{InfoCA-ms}. This was seen clearly with certain class II rules, where at one scale they have a high $E$, but a minimal at higher scales. The change of the measures across scales provides a useful tool to study dynamical systems~\citep{Wolpert:1999,Polani:2004,Weeks:2010}, as it has been formalized by Krohn-Rhodes theory~\citep{Rhodes:2009,Egri-Nagy:2008}. For example, a system with a high (but not maximum) $E$ across scales can be said to be more emergent than one with $E$ close to maximum at one scal e but close to minimal at other scales.}

\textcolor{black}{Combinations of the measures can be used to better understand and classify systems. Looking only at $C$ or $H$, it is not possible to distinguish complex from chaotic rules, but their combination reveals the ether required for the interaction of complex structures. There is also the question of \emph{how} to consider and measure the information used by the measures. For example, in ECA different results are obtained if bit strings are considered vertically (as we did), horizontally, or diagonally.
}

The results of RBNs change little with scale. This is precisely because RBNs are randomly generated. ECA are much more regular (in structure and function). Still, there are ECA which do not change much with scale, either they are too ordered (class I) or too chaotic (some of class III). In between, there are several rules with characteristic patterns which make scale matter. Most phenomena do have characteristic patterns, so their scale should also be relevant.

For RBNs and ECA with chaotic dynamics, emergence is reduced slightly with increasing scale. This is a finite size effect, because \textcolor{black}{shorter} strings have their emergence reduced more than \textcolor{black}{longer} ones. This is interesting, since many studies on dynamical systems, especially analytic, are made assuming infinite systems. However, most real systems are better modeled as finite, and here we can see that the length of a string---not only the scale---can play a relevant role on determining the emergence and complexity of systems. One implication of this finite size effect is that higher scales require less information to be described. Extrapolating, the ``highest scale" ($b \rightarrow \infty$) implies no information ($I \rightarrow 0$): if everything is contained, then there is no information \textcolor{black}{necessary} to describe it~\citep{Gershenson:2007}.

\section{Future Work}

There are several lines of future research that we would like to pursue:

\begin{enumerate}

\item \textcolor{black}{We are applying the proposed measures to ultra-large scale systems~\citep{Northrop:2006} with interesting preliminary results~\citep{Amoretti:2012}.}

\item \textcolor{black}{We intend to apply the proposed measures to time series from real systems, as well as considering a non-random $I_{in}$. This will enable us to contrast the usefulness of the measures.}

\item We studied measures for RBNs with computer simulations. It would be relevant to develop analytical solutions and contrast these to our simulation results.

\item There are some similarities between $C$ and Fisher information~\citep{Prokopenko:2011,wang2011fisher}. It would be \textcolor{black}{useful} to study their potential relationship. 

\item \textcolor{black}{Bennet's logical depth~\citep{Bennet:1995} has been recently used as a measure of image complexity~\citep{Zenil:2012}. The relationships between logical depth and the proposed measure of complexity are worth studying.}

\item We would like to evaluate our proposed measures on Turing machines~\citep{Delahaye:2007,Delahaye:2012} and on $\epsilon$-machines~\citep{ShaliziCrutchfield2001,GoernerupCrutchfield2008}. \textcolor{black}{The latter ones will allow for a direct comparison of our proposed measures with the ones of Crutchfield, Shalizi, et al.~\citep{Crutchfield:1989,Shalizi2001,Shalizi:2004}}.

\item \textcolor{black}{Shannon's information was recently used as a measure of spatial complexity, with applications to urban areas~\citep{Batty:2012}. It would be interesting to compare this work with our proposed measure of complexity.}

\item We are interested in studying a new complexity profile, combining Bar-Yam's profile and the $\sigma$ profile to study the complexity of different phenomena at different scales.

\item The multiple scale approach could be extended to include the \emph{meaning} of information~\citep{Neuman:2008}. Just like the same string can change its complexity with scale, information can change its meaning with context~\citep{Edmonds2001,Gershenson2002ua,Edmonds:2012}.

\item We can extend our approach to measure and study autopoiesis~\citep{VarelaEtAl1974,MaturanaVarela1980,McMullin2004}, using the concept of life ratio~\citep{Gershenson:2007}:
\begin{equation}
A=\frac{I_{self}}{I_{env}}
\label{eq:A}
\end{equation}
where the $A$ is the ratio between the information produced by a system over the information produced by its environment.

\item Our measures could be generalized to computing networks~\citep{Gershenson:2010b}, with potential applications \textcolor{black}{to} the study complex networks~\citep{Barabasi2002,Newman:2003,NewmanEtAl2006,Caldarelli:2007,Newman:2010}.

\end{enumerate}

So many research questions are beyond the capabilities of the authors (another finite size effect). We invite the community to explore these questions in collaboration or independently.

\section{Conclusions}

We proposed abstract measures of emergence, self-organization, complexity, and homeostasis based on information theory with the intention of clarifying their meaning with formal definitions. We illustrated the measures with computational experiments of discrete dynamical systems (RBNs and ECA).

Emergence can be defined as a measure of the information that a system produces, relative to the information that the system receives. Self-organization can be defined as the difference between the information the system receives (input) and the information the system produces (output). These measures can be simplified assuming random inputs, in which case emergence becomes the opposite of self-organization. 
Complexity represents the balance between self-organization and emergence, related to the balance of order and chaos. There are different methods to guide dynamical systems to this balance~\citep{Gershenson:2010}. This balance can be achieved by natural selection, or also by designers aiming to exploit the benefits of complexity. Homeostasis  measures the stability of information in time. Our measures can be applied at different scales, yielding different results in particular cases. There are several research topics in which the ideas presented here could be extended.

Complexity, emergence, self-organization, and homeostasis are present all around us (everything can be described in terms of information~\citep{Gershenson:2007}, so we can measure the $C$, $E$, $S$, and $H$ of everything). Given their pervasiveness, we should all make an effort to further understand and study these properties of all phenomena.

\section*{Acknowledgments}

We should like to\ thank \textcolor{black}{Michele Amoretti, Mario Cosenza, Carlos Maldonado, Daniel Polani, Mikhail Prokopenko, H\'ector Zenil, and two anonymous referees for fruitful exchanges and useful suggestions}. This work was supported SNI membership 47907 of CONACyT, Mexico, UNAM-DGAPA-IACOD project T2100111, by Intel\textsuperscript{\textregistered}, and by IIMAS, UNAM with a visiting professor fellowship granted to N.F.

\clearpage

\bibliographystyle{cgg}

\bibliography{carlos,sos,RBN,complex,information,COG}

\end{document}